\documentclass[10pt,aps,prd,twocolumn,floats,floatfix,showpacs,superscriptaddress,nofootinbib]{revtex4-1}

\usepackage[utf8]{inputenc}
\usepackage[T1]{fontenc}

\usepackage{enumerate}

\usepackage[usenames]{xcolor}
\usepackage{amsmath,amssymb}
\usepackage[colorlinks]{hyperref}
\usepackage{cleveref}

\newcommand{\note}[1]{\text{\tiny{#1}}}

\newcommand{\dd}{\mathrm{d}}
\newcommand*\DAlembert{\mathop{}\!\mathbin\Box}
\renewcommand{\rho}{\varrho}
\renewcommand{\epsilon}{\varepsilon}

\def\beq{\begin{equation}}
\def\eeq{\end{equation}}
\def\bea{\begin{eqnarray}}
\def\eea{\end{eqnarray}}

\def\L{\mathcal{L}}

\def\H{\mathcal{H}}
\usepackage{mathtools}
\usepackage[english]{babel}

\newcommand\eqh{\stackrel{\H}{=}}
\newcommand\equivh{\stackrel{\H}{\equiv}}
\newcommand\proph{\stackrel{\H}{\propto}}

\begin{document}

\title{Causal structure of black holes in shift-symmetric Horndeski theories}

\author{Robert Benkel}
\affiliation{School of Mathematical Sciences, University of Nottingham,
University Park, Nottingham NG7 2RD, United Kingdom}

\author{Nicola Franchini}
\affiliation{School of Mathematical Sciences, University of Nottingham,
University Park, Nottingham NG7 2RD, United Kingdom}

\author{Mehdi Saravani}
\affiliation{School of Mathematical Sciences, University of Nottingham,
University Park, Nottingham NG7 2RD, United Kingdom}

\author{Thomas P.~Sotiriou}
\affiliation{School of Mathematical Sciences, University of Nottingham,
University Park, Nottingham NG7 2RD, United Kingdom}
\affiliation{School of Physics and Astronomy, University of Nottingham,
University Park, Nottingham NG7 2RD, United Kingdom}
\affiliation{Centro de Astrofisíca e Gravitação--CENTRA, Instituto Superior Técnico--IST, Universidade de Lisboa--UL, Avenida Rovisco Pais 1, Lisboa 1049-001, Portugal}

\begin{abstract}
In theories with derivative (self-)interactions, the propagation of perturbations on nontrivial field configurations is determined by effective metrics. Generalized scalar-tensor theories belong in this class and this implies that the matter fields and gravitational perturbations do not necessarily experience the same causal structure. Motivated by this, we explore the causal structure of black holes as perceived by  scalar fields in the Horndeski class. We consider linearized perturbations on a {\em fixed} background metric that describes a generic black hole. The effective metric that determines the propagation of these perturbations does not generally coincide with the background metric (to which matter fields couple minimally). Assuming that the metric and the scalar respect stationarity and that the surface gravity of the horizon is constant,  we prove that Killing horizons of the background metric are always Killing horizons of the effective metric as well. Hence, scalar perturbations cannot escape the region that matter fields perceive as the interior of the black hole. This result does not depend on asymptotics but only on local considerations and does not make any reference to no-hair theorems. We then demonstrate that, when one relaxes the stationarity assumption for the scalar, solutions where the horizons of the effective and the background metrics do not match can be found in the decoupling limit.
\end{abstract}

\maketitle

\section{Introduction}

Black holes are among the simplest objects in the Universe and general relativity (GR) predicts that their defining feature, the event horizon, acts as a causal boundary for all fields. Black holes are the perfect probes of the strong-field regime, especially in the era of gravitational wave astronomy. This regime is also where alternative theories of gravity are expected to yield new predictions in the description of compact astrophysical objects. When new fundamental fields become a part of the picture, the simplicity of black holes in GR, as expressed by the no-hair conjecture \cite{Israel:1967wq,Carter:1971zc,Robinson:1975bv}, is often replaced by solutions with a more complex structure.

Even the simplest extensions of GR, scalar-tensor theories which postulate that a scalar field takes part in mediating the gravitational interaction, can introduce significant deviations \cite{Sotiriou:2015lxa,Sotiriou:2015pka}. Consider  the most general scalar-tensor theory that leads to second order equations upon direct variation, described by the Horndeski action \cite{Horndeski:1974wa,Deffayet:2009mn,Kobayashi:2011nu}
\begin{equation}
\label{eq:horndeski_lagrangian}
	\L_\note{Horndeski}
	=
	\L_{2} + \L_{3} + \L_{4} + \L_{5}
\end{equation}
where
\begin{align}
\label{eq:horndeski_lagrangian_terms}
	\L_{2} &= G_{2}\,, \\
	\L_{3} &= - G_{3} \Box \Phi\,, \\
	\L_{4} &= G_{4} R + G_{4X} \left[ \left( \DAlembert \Phi \right)^{2} - \left( \nabla_{a} \nabla_{b} \Phi \right)^{2} \right]\,, \\
	\L_{5} &= G_{5} G_{ab} \nabla^{a} \nabla^{b} \Phi \\ &- \frac{1}{6} G_{5X} \left[ \left( \DAlembert \Phi \right)^{3} - 3 \DAlembert \Phi \left( \nabla_{a} \nabla_{b} \Phi \right)^{2} + 2 \left( \nabla_a \nabla_b \Phi \right)^{3} \right]\,,
\end{align}
$X = - \frac{1}{2} \nabla_{a} \Phi \nabla^{a} \Phi$, $G_{i}=G_{i}(\Phi,X)$, and $G_{iX}=\partial_{X} G_{i}$.
No-hair theorems do exist for broad subclasses of the Horndeski action under certain assumptions:  when the scalar does not exhibit derivative self-couplings, $G_{2}= G_{2}(\Phi)$, $G_{4}= G_{4}(\Phi)$, $G_3=G_5=0$, and the configuration is stationary and asymptotically flat \cite{1970CMaPh..19..276C, Hawking:1972qk,Bekenstein:1995un,Sotiriou:2011dz}; when the theory respects shift symmetry, $\Phi \to \Phi+$constant, $G_{i}= G_{i}(X)$,  assuming staticity and spherical symmetry \cite{Hui:2012qt} or slow rotation \cite{Sotiriou:2013qea}, and provided that the scalar does not couple to the Gauss-Bonnet invariant, $\mathcal{G}=R_{abcd}R^{abcd}-4R_{ab}R^{ab}+R^2$ \cite{Sotiriou:2013qea,Sotiriou:2014pfa}. However, there is no known no-hair theorem that covers the complete action. On the contrary, it is known to admit hairy solutions even for simple static, spherically symmetric and asymptotically flat configurations, {\em e.g.}~\cite{Kanti:1995vq,Yunes:2011we,Sotiriou:2013qea,Sotiriou:2014pfa,Silva:2017uqg,Doneva:2017bvd,Antoniou:2017acq}. Moreover, in certain cases it is possible to obtain stationary hairy black hole spacetimes by relaxing the stationarity assumption for the scalar only \cite{Herdeiro:2014goa,Babichev:2013cya,Charmousis:2014zaa,Kobayashi:2014eva}.

The existence of hairy black hole solutions in theories where the scalar exhibits derivative self-interactions might have implications for causality. Noncanonical kinetic terms allow for perturbations of the scalar field to propagate superluminally as long as the background field is nontrivial, i.e. has nonvanishing derivatives \cite{Babichev:2007dw,Babichev:2017lmw,Babichev:2018uiw}. Indeed, as a simple approximation, consider the Lagrangian \eqref{eq:horndeski_lagrangian} as describing a scalar field on a {\em fixed} curved background. Expanding to second order in scalar perturbations $\pi$ on top of a background $\varphi$, we get the second order Lagrangian for scalar perturbations as follows
\begin{equation}\label{eq:effective metric definition}
	\L^{(2)}_\note{Horndeski}
	=-\frac{1}{2}f^{ab}[\varphi,g] \partial_{a} \pi \partial_{b} \pi.
\end{equation}
The explicit form of $f^{ab}$ is presented in Appendix \ref{appendix_effective_metric} for the special case where $\Phi$ respects shift symmetry. The propagation of linear scalar perturbations is determined by effective metric $f^{ab}$. Hence, this metric, or its inverse $\left(f^{-1}\right)_{ab}$, where $f^{ab}\left(f^{-1}\right)_{bc}=\delta^a_c$, define causality for such excitations. The matter fields are assumed to couple minimally to $g_{ab}$ and hence massless matter perturbations will follow null trajectories of that metric. For a generic background scalar field $\varphi$, $f^{-1}_{ab}$ does not coincide with  $g_{ab}$.  This suggests that the causal structure of black holes in certain classes of theories described by the Horndeski action should be particularly intriguing.  If $f^{-1}_{ab}$ and $g_{ab}$ are not conformally related then they have different null cones, so massless excitations that follow null geodesics of $f^{-1}_{ab}$ can be superluminal or subluminal.

One can take this a step further: for stationary metrics, event horizons will be Killing horizons, so one can ask under which conditions the Killing horizons of $f^{-1}_{ab}$ and $g_{ab}$ coincide and what is the causal structure when they do not. These are precisely the questions that we explore below. We restrict ourselves to a setup where the background metric is fixed and consider only linear scalar perturbations, as in the discussion above. This has obvious limitations: it only probes the local counterpart of the causal structure as perceived by scalar excitations and ignores nonlinear effects. It also does not take into account the role of metric perturbations, which will generally also propagate along null geodesics of a different effective metric. Nonetheless, this setup makes calculation tractable and already provides very interesting insights into the causal structure as perceived by the scalar.

The causal structure of black holes in Horndeski theories has been studied before in \cite{Tanahashi:2017kgn} and \cite{Minamitsuji:2015nca} (see \cite{Reall:2014pwa} and \cite{Izumi:2014loa} for similar approaches) using the method of characteristics. In Ref.~\cite{Tanahashi:2017kgn} both the metric  $g_{ab}$ and the scalar field $\varphi$ were taken to be dynamical and it was claimed that if a surface $\Sigma$ is a Killing horizon of  $g_{ab}$, it is also a characteristic surface for all degrees of freedom when the metric and the scalar are stationary. It was then suggested that this implies that $\Sigma$ effectively acts as a horizon for all excitations. This last statement entails some implicit assumptions not spelled out in Ref.~\cite{Tanahashi:2017kgn}. Our approach follows a different path and it provides a rigorous proof about when Killing horizons for different excitations coincide. Though less general due to the decoupling approximation, it is also more transparent physically. Hence, it simplifies the interpretation of the results and highlights certain subtleties and assumptions in the characteristics approach.

The paper is organized as follows:  in the next section, we provide a proof that the Killing horizons of the fixed background metric $g_{ab}$ will also be Killing horizons of $f^{-1}_{ab}$, provided that both the metric and the scalar respect stationarity. In Sec.~\ref{sec:t-dep} we relax the assumption of stationarity for the scalar and we show that, in the zero backreaction limit, it is rather straightforward to construct solutions where the horizons of these two metrics do not coincide. Though it is not clear if this feature will survive once backreaction is taken into account, our results certainly motivate further work in this direction. Sec.~\ref{sec:discuss} contains a discussion of our results and their potential implications.

\section{No-go results for black holes with multiple horizons}

As argued above $f^{-1}_{ab}$ and $g_{ab}$ do not coincide for a generic background. In fact, interestingly, they do not coincide even for the trivial background, $\varphi=$ constant. However, we will prove below that the black hole Killing horizon $\H$ perceived by the metric $g_{ab}$ acts as a Killing horizon for $f^{-1}_{ab}$ metric as well. In order to do so, we assume that the black hole spacetime has a Killing vector field $\xi^a$ which is timelike outside the black hole region, is null and orthogonal to $\H$ on the horizon $\H$ and the background scalar field satisfies this symmetry, $\xi^a \nabla_a \varphi = 0$.
Moreover, we assume those derivatives of $G_i$'s appearing in the expression for $f^{ab}$ in Appendix \ref{appendix_effective_metric} are finite.

Let us start with the easiest case, where the scalar background is trivial, $\varphi=$ constant. The effective metric for linear perturbations around this solution is given by (see Appendix \ref{appendix_effective_metric})
\beq\label{effective_metric_vacuum}
f^{ab}=c_1 g^{ab}+c_2 G^{ab}
\eeq
where $c_1$ and $c_2$ are constants.
Consider a spacetime where no extra matter field is present. Then, the variation of the action w.r.t. $g^{ab}$ yields the vacuum Einstein equations $G_{ab}+\lambda g_{ab}=0$. As a result, $f^{ab}$ and $g^{ab}$ are related by a conformal factor and their causal structure is the same.

In general, when the scalar field background is trivial, the $g_{ab}$ metric satisfies GR field equations, and thus describes a GR black hole. The most general GR black hole solution is described by a Kerr-Newman metric. Though not obvious,  we have verified by direct calculation that the Killing horizon of the Kerr-Newman black hole acts as a horizon for the $f^{-1}_{ab}$ metric as well.\footnote{Presumably this statement can be proven using the properties of the electromagnetic field, but we have not attempted that as it is covered by the more general proof given below.} In the argument above, we have exploited the field equation for the metric to derive our result. In what follows, we provide a proof to show that the Killing horizon of $g_{ab}$ is a Killing horizon of the effective metric $f^{-1}_{ab}$  without any reference to the field equation for  $g_{ab}$ (or to the equation for the background scalar field $\varphi$).


 $f^{-1}_{ab}$ is a composite metric constructed from the spacetime metric and the scalar field derivatives. Consequently, it satisfies stationarity, i.e. $\L_{\xi}f^{-1}_{ab}=0$ where $\L_\xi$ is the Lie derivative along $\xi^a$.
The Killing horizon $\H$ is a Killing horizon for the effective metric, iff
\bea
\mbox{I}:&& ~f^{ab}\xi_a\xi_b \eqh 0, \label{cond1}\\
\mbox{II}:&& ~f^{ab}\xi_a \eqh \aleph \xi^b, \label{cond2}
\eea
where $\eqh$ means equality holds on $\H$ and $\aleph$ is a scalar.
We ignore $\aleph = 0$ which corresponds to a degenerate perturbation metric, signalling the instability of background scalar configuration.
Condition I states that the Killing horizon $\H$ is a null surface w.r.t.~the effective metric. Condition II requires that the Killing vector field  be hypersruface orthogonal on $\H$ - a necessary condition for the hypersurface to be a Killing horizon.  Indeed, multiplying both sides of Eq. \eqref{cond2} with $f^{-1}_{ab}$, we get
\beq
\label{interm}
\xi_a \eqh \aleph f^{-1}_{ab} \xi^b.
\eeq
$\xi_a$ is the normal to $\H$, thus this equation implies that $f^{-1}_{ab}\xi^b$ is normal to $\H$.

In Appendix \ref{identity_proof}, we prove both conditions. For condition I we only need to assume that the background metric and scalar configuration satisfy stationarity. To prove condition II without having to resort to global considerations, we need to make the additional assumption that
 {\em  the surface gravity of the horizon is constant}. Overall, we have proved that the spacetime metric horizon $\H$ is a Killing horizon for the perturbation metric. Hence scalar perturbations cannot escape the region that matter fields perceive as a black hole. Note that if the scalar perturbations are subluminal, it is conceivable that $f^{-1}_{ab}$ could have another horizon outside $\H$.

Clearly, condition II implies condition I.  Multiplying both sides of the  Eq.~\eqref{interm} with $\xi^a$ yields
\beq\label{app:fhorizon}
f^{-1}_{ab}\xi^a \xi^b \eqh 0
\eeq
which means $\xi^a$ is null w.r.t. the perturbation metric on $\H$. We have chosen to consider it as separate condition to make better contact with the results of Ref.~ \cite{Tanahashi:2017kgn}. Condition I also identifies the characteristics for scalar perturbations, as is discussed in more detail in Appendix \ref{app:ref_comparison}. As shown there, if one applies the approach of Ref.~ \cite{Tanahashi:2017kgn} to the decoupling limit  and considers perturbation around a stationary background configuration, one does indeed obtain condition I.

 The characteristics can be thought of as maximal speed of propagation surfaces and the boundary of the future of a {\em given} subset of the manifold is a characteristic. Nonetheless, characteristics are clearly not horizons in general (any null surface in Minkowski space is a characteristic for the wave equation) and condition I does not imply condition II without further assumptions. So, proving the former condition is not sufficient to claim that the Killing horizon of $g_{ab}$ will also be a Killing horizon of $f^{-1}_{ab}$ given that the two metrics are distinct. One can argue around this technical obstruction. Condition I does show that scalar excitations can cross the Killing horizon of $g_{ab}$ in a single direction only, which depends on the orientation of the characteristic. Hence, assuming that both $g_{ab}$ and $f^{-1}_{ab}$ are  asymptotically flat and identifying their asymptotic regions should suffice to claim that scalar perturbations are trapped by the Killing horizon of $g_{ab}$ and cannot reach infinity, as claimed in Ref.~ \cite{Tanahashi:2017kgn}.

 It is hard to imagine a physical solution that would not satisfy these assumptions and yet it is interesting that they are  not sufficient to prove that $\H$ is a Killing horizon of $f^{-1}_{ab}$ without further assuming that the surface gravity is constant. Carter has proven without resorting to field equations that Killing horizons in axisymmetric spacetimes that satisfy ``$t$--$\phi$ orthogonality'' have constant surface gravity  \cite{Carter:1973rla}.\footnote{By ``$t$--$\phi$ orthogonality'' one refers to the requirement that the $t$--$\phi$ plane be orthogonal to a family of 2-dimensional surfaces, which is a prerequisite for the line element to have $g_{t\phi}$ as the only off-diagonal component.} The  is also trivially true in spherical symmetry. Indeed, in what follows we present a complementary proof that is valid in spherical symmetry and does not make use of this assumption.

 It is also worth mentioning that, if one is willing to assume that the Killing vector is timelike w.r.t. $f^{-1}_{ab}$ in the exterior of $\H$, then there seems to be an alternative way to argue that condition II has to hold. The Killing flow cannot pierce  $\H$, as the latter is a Killing horizon for $g_{ab}$. Hence, $\xi^a$  either has to be orthogonal to $\H$ and null w.r.t. $f^{-1}_{ab}$ as well, or is has to reside in $\H$ and be spacelike w.r.t.  $f^{-1}_{ab}$. However, the latter case is excluded by continuity if $\xi^a$ is timelike immediately outside $\H$.

\subsection*{Special case: Spherical symmetry}

For a static spherically symmetric spacetime and scalar field configuration, we make use of the $\{t,r,\theta,\phi\}$ coordinates. In this coordinate system, both spacetime and effective metrics are diagonal and given by
\bea
g_{ab} dx^a dx^b&=& - H(r) dt^2 + \frac{dr^2}{F(r)} + r^2 d\Omega^2,\label{eq:metric}\\
(f^{-1})_{ab} dx^a dx^b&=&- \tilde H(r) dt^2 + \frac{dr^2}{\tilde F(r)} + \beta(r)^2r^2 d\Omega^2,\qquad
\eea
where $d\Omega^2=d\theta^2 + \sin\theta^2 d\phi^2$. The Killing vector is given by $\xi^a=(1,0,0,0)$ and the Killing horizon of the effective metric is where $f^{-1}_{tt}=-\tilde H(r)=0$, or equivalently $ f^{tt}=\infty$.

The Killing horizon of the effective metric cannot be located away from the metric Killing horizon at $r=r_H$ where $H(r_H)=F(r_H)=0$. The argument goes as follows: away from the metric Killing horizon, all components of the spacetime metric and its inverse, in $\{t,r,\theta,\phi\}$ coordinates, are finite. This means that all components of the Riemann tensor and field derivatives (with upper or lower indices) are finite. $f^{tt}$ is given by different combinations of the Riemann tensor and field derivatives contractions (see Appendix \ref{appendix_effective_metric}), and thus it is finite and cannot satisfy the condition for the effective metric Killing horizon $f^{tt}=\infty$.

\section{Time-dependent scalars and black holes with multiple horizons}
\label{sec:t-dep}

The theorem in the previous section ruled out the possibility of having multiple horizons for most stationary configurations in a large class of shift-symmetric Horndeski theories. However, as mentioned in the Introduction,  static, spherically symmetric metric with time-dependent scalar configurations are known to exist in shift-symmetric Horndeski theories \cite{Babichev:2013cya,Charmousis:2014zaa,Kobayashi:2014eva}. In particular, if $\Phi$ depends linearly on Killing time $t$, {\em i.e.} $\Phi= q t+\psi(r)$ in a suitable coordinate system and $q$ is a constant, then its derivatives respect staticity. Shift symmetry implies that $\Phi$ is present in the equations only through its derivatives, and hence the mismatch between the symmetries of the metric and the scalar does not necessarily lead to an inconsistency.
This motivates the study of causality and the potential existence of multiple horizons in solutions with stationary metrics and time-dependent scalars. We will not attempt an exhaustive analysis here. Instead, we will just provide a simple example of a solution in which the Killing horizons of $g_{ab}$ and $f^{-1}_{ab}$ do not match, as evidence that this issue deserves further investigation.

One might actually expect the solutions of Refs.~\cite{Babichev:2013cya,Charmousis:2014zaa} to readily provide such an example. Unfortunately, this is not the case within the confines of the zero-backreaction limit that we are employing here.
This becomes apparent when one briefly examines the hairy black hole solutions  presented in \cite{Babichev:2013cya}.
The action used there is a combination of $\L_2$ and $\L_4$
\begin{equation}\label{eq:actionbabichev}
	\L
	=
	\zeta R-2\eta X+\beta G^{ab}\nabla_a\Phi\nabla_b\Phi-2\Lambda
	\,,
\end{equation}
which corresponds to $G_{2}=2 \eta X - 2 \Lambda$ and $G_{4}=\zeta + \beta X$.
Shift symmetry implies that the equation for the scalar field can be written as a current conservation equation
\begin{equation}
	\nabla_a J^a
	=
	0
	\,,
\end{equation}
where
\begin{equation}
\label{eq:noether_current}
	J^{a}
	=
	(\eta g^{ab} - \beta G^{ab})\partial_b\Phi
	\,.
\end{equation}

The field equation for $\Phi$ is linear in $\Phi$. Hence, one considers scalar perturbations in a {\em fixed} hairy background, consistent with our assumptions in the previous sections, $f^{ab}=2\eta g^{ab} - 2\beta G^{ab}$. If that background is a solution of vacuum Einstein's equation with or without a cosmological constant, the $f_{ab}$ and $g_{ab}$ are conformally related and share their horizons.

 Let us instead assume that $g_{ab}$ is one of the static, spherically symmetric metrics of the hairy solution reported in Ref.~ \cite{Babichev:2013cya}. The following ans\"{a}tze were used
\begin{align}
	ds^2
	&=
	- h(r) dt^2 + \frac{dr^2}{f(r)} + r^2 d\Omega^2
	\,, \\
	\Phi(t,r)
	&=
	q t + \psi(r)
	\,,
\end{align}
and all of the solutions found turn out to satisfy
\begin{equation} \label{eq:rr_condition}
	 \eta g^{rr} - \beta G^{rr}
	=
	0
\end{equation}
everywhere in the spacetime. However, Eq.~\eqref{eq:rr_condition} implies that  $f^{rr}=0$ everywhere and, hence, the effective metric for linear perturbations is degenerate throughout spacetime when backreaction is neglected. Clearly, in such solutions one cannot discuss in a meaningful way  Killing horizons of $f^{ab}$ in the zero-backreaction limit. The degeneracy of $f^{ab}$ is clearly related to the choice of theory and the expression for the effective metric. Nonetheless, it does not actually imply that there is some  general pathology of the solutions or the theory, as it appears only when one ignores the metric perturbations but still uses a background with a nontrivial scalar ({\em i.e.}~when backreaction is neglected selectively at first order only.) See Refs.~\cite{Ogawa:2015pea,Babichev:2017lmw,Babichev:2018uiw} for discussions on the linear stability of these solutions.

In order to provide an example of a solution with multiple horizons without taking into account backreaction, we turn to a different Lagrangian,  known as k-essence,
\begin{equation}
\label{eq:kessence}
	\L
	=
	G_2[X]
	\,,
\end{equation}
where $G_{2X}>0$. Varying this action with respect to $\Phi$, we arrive at the equation of motion
\begin{equation}
\label{eq:kessence_scalar}
	\left(G_{2X} g^{ab}-G_{2XX}\nabla^a\Phi \nabla^b\Phi\right)\nabla_a\nabla_b\Phi
	=
	0
	\,.
\end{equation}
 The causal structure of linear perturbations $\pi$ on top of a background field $\varphi$ is governed by
\begin{equation}\label{K_essence_metric_upper}
f^{ab}=G_{2X}g^{ab}-G_{2XX}\nabla^a\varphi\nabla^b\varphi.
\end{equation}
Inverting \eqref{K_essence_metric_upper} yields
\begin{equation}\label{K_essence_effective_metric}
\left(f^{-1}\right)_{ab}=\frac{1}{G_{2X}}\left(g_{ab}+\frac{G_{2XX}}{G_{2X}+2XG_{2XX}}\nabla_a\varphi\nabla_b\varphi\right)
\end{equation}
which determines how scalar perturbations propagate through the spacetime.

Let us assume that $g_{ab}$ describes a static spherically symmetric black hole spacetime with time-translational Killing vector $\xi^a$. The Killing horizon is located where $\xi^a$ becomes null according to the spacetime metric, i.e.
\begin{equation}
	g_{ab} \xi^a \xi^b
	=
	0
	\,.
\end{equation}
Scalar perturbations can evade the metric horizon provided that $\xi^a$ remains timelike with respect to the effective metric $(f^{-1})_{ab}$ on the Killing horizon. This happens when
\begin{align}
	&\xi^a \nabla_a \varphi
	\neq
	0 \label{eq:phi_time_dependent}
	\, \\
	&\frac{G_{2XX}}{G_{2X}+2XG_{2XX}}
	<
	0
\end{align}
on the metric horizon. The first condition expresses that the scalar field is time dependent and the second one means that the perturbations are superluminal. It is important to note that the above conditions are local, i.e. they need to be satisfied only in a neighborhood of the Killing horizon.

We now proceed to construct an explicit solution for the scalar field in a {\em Schwarzschild background} with mass $M$, and demonstrate that the horizon $f^{-1}_{ab}$ does not coincide with that usual horizon.  To do so, we make a specific choice of k-essence theory \eqref{eq:kessence}, with
\begin{equation}
	G_2
	=
	X + \frac{1}{2} \alpha X^2
	\,,
\end{equation}
where $\alpha>0$ ($\alpha<0$) corresponds to subluminal (superluminal) propagation. According to Eq. \eqref{eq:phi_time_dependent}, time dependence in $\varphi$ is necessary to have multiple horizons.  We choose the following ansatz
\begin{equation}
	\varphi(v,r)
	=
	q v + \psi(r),
\end{equation}
where $v=t+r_*$ is the ingoing Eddington-Finkelstein coordinate, and $r_*$ is the tortoise coordinate defined as $\dd r_*=\frac{\dd r}{1-2M/r}$. We further assume that our solution is continuously connected to $\varphi=$constant as $q\to0$, and we expand $\psi(r)$ as follows,
\begin{equation}
	\psi(r)
	=
	\sum_{n=1}^{\infty} \psi_n q^n
	\,.
\end{equation}
We then solve the equations perturbatively in $q$.
At first order in $q$, Eq. \eqref{eq:kessence_scalar} reads
\begin{equation}
	\frac{2(M-r)\psi_1'-r[2+(r-2M)\psi_1'']}{r^2}=0,
\end{equation}
whose solution, after imposing regularity on the horizon, is
\begin{equation}
	\psi_1'(r)=-\left(1+\frac{2M}{r}\right).
\end{equation}
With this solution and ignoring the conformal factor $G_{2X}$ which is irrelevant for the causal structure, the effective metric \eqref{K_essence_effective_metric} yields
\begin{multline}\label{eq:effmetricpert}
\left(f^{-1}\right)_{ab}dx^adx^b \propto\left(\alpha q^2 -1+\frac{2M}{r}\right)dv^2 \\
+2\left[1-\alpha q^2\left(1+\frac{2M}{r}\right)\right]drdv \\
+\alpha q^2 \left(1+\frac{2M}{r}\right)^2 dr^2 +r^2 d\Omega^2.
\end{multline}
We want to express metric \eqref{eq:effmetricpert} in Eddington-Finkelstein-like coordinates,
thus we perform the following coordinate transformation
\bea
\dd v = \dd v' \left(1+\frac{1}{2}\alpha q^2\right)-\dd r\frac{\alpha q^2}{2}\left(1+\frac{2M}{r}\right)^2.
\eea
Defining $M' = M\left(1+\alpha q^2\right)$, up to second order in $q$ the metric will read
\begin{multline}
\left(f^{-1}\right)_{ab}\dd x^a\dd x^b\propto-\left(1-\frac{2M'}{r}\right)\dd v'^2 \\
+2\left[1-\frac{\alpha q^2 M'(4M'^2+2M' r+r^2)}{r^3}\right]\dd r\dd v'+r^2\dd\Omega^2.
\end{multline}

Starting from $r>2M'=2M(1+\alpha q^2)$, outgoing radial null rays of the effective metric can escape to infinity. For $\alpha<0$, corresponding to the superluminal case, this region extends to both sides of the Killing horizon of the background metric $g_{ab}$.

Our solution clearly provides an example of an effective metric whose horizon does not coincide with that of the background metric, albeit is a much simplified setup where there is no backreaction and where the background is that of GR. It can be seen as having been obtained at the decoupling limit, where there is no backreaction at all orders in perturbation theory. The potential caveats are: (i) once backreaction is taken into account, the causal structure of the effective metric can change; (ii) it is not guaranteed that beyond decoupling, a static metric with a time-dependent scalar actually exists. On the other hand, one can argue that if the scalar field configuration evolves very slowly in time compared to the characteristic timescale defined by the mass of the black hole, then treating the black hole as quasistationary might be a good approximation that justifies our treatment. In any case, our sole purpose here is to demonstrate that the possibility of having multiple horizons deserves further investigation when it comes to time-dependent scalars.

\section{Discussion}\label{sec:discussion}
\label{sec:discuss}
The causal structure of black holes in Horndeski theories can deviate from the causal structure dictated by the spacetime metric. Due to derivative couplings of the scalar field and metric, the perturbation metrics are not conformal to the spacetime metric. Here we have focused for simplicity on the causal structure as seen by the linear perturbations of the scalar field on a {\em fixed} background metric.

The main result of this paper is the following: although null cones of the linear perturbation metric ($f^{-1}_{ab}$) and the spacetime metric ($g_{ab}$) are generally different, we have shown that for stationary black holes and scalar fields, and provided that the surface gravity of the horizon is constant, a Killing horizon of $g_{ab}$ $\H$ is also a Killing horizon for $f^{-1}_{ab}$.  This means that the black hole region as perceived by matter fields, which are minimally coupled to $g_{ab}$, is a subset of or the same as the one defined by linear perturbations of the scalar field.

Our results agree with previous results in the literature that used the method of characteristics \cite{Tanahashi:2017kgn} when there is overlap, but they go a bit further to prove that $\H$ is actually a Killing horizon for $f^{-1}_{ab}$. They also help interpret previous result in the intuitive physical picture of effective metrics and pin down certain subtleties of the characteristics approach. Remarkably, to rigorously complete the proof that $\H$ is a Killing horizon for $f^{-1}_{ab}$ for a generic stationary configuration one  needs the local assumption of constant surface gravity. Alternatively, one can argue that $\H$ is a causal boundary of all excitations simply by being a characteristic, and hence is expected to be a Killing horizon for $f^{-1}_{ab}$. This correspondence of assumptions begs the conjecture that Killing horizons have constant surface gravity for  stationary, asymptotically flat black hole configurations in Horndeski theories. Though there is no known proof of that statement (or a theory-independent proof), it is worth mentioning the following. Spherical black holes always possess this property and Carter has proven that axisymmetric black holes will also have this property if they satisfy ``$t$--$\phi$ orthogonality'' \cite{Carter:1973rla}. The latter is a necessary condition for having a line element with $g_{t\phi}$ as the only off-diagonal component (the usual ansatz for stationary, axisymmetric solutions). Finally, the zeroth law of black hole thermodynamics is related to the fact that surface gravity is constant.
Hence, finding black hole solutions that do not satisfy this property would indeed be remarkable.

It is important to clarify the complementarity between our proof and no-hair theorems. In our general proof we do not impose that the background configuration satisfies any specific field equations and, hence, we did not need to make any reference to no-hair theorems as it applies equally well to hairy solutions. As has been discussed above, for backgrounds with a trivial scalar configuration the proof becomes rather trivial. This might give the false impression that the physical significance of our result is limited. To clarify this point, we would like to emphasize that stationary hairy solutions do exist in shift-symmetric Horndeski theories and that our proof does not rely on most of the assumptions employed in no-hair theorems. In particular, we bring the following points to the reader's attention:
\begin{enumerate}[(i)]
\item Our proof uses only local properties of Killing horizons and makes no reference to the asymptotics. Hence, it applies to black holes with nonflat asymptotics or matter in their vicinity, which are not covered by no-hair theorems.
\item The known no-hair theorems for shift-symmetric Horndeski theories require the assumptions of staticity and spherical symmetry \cite{Hui:2012qt} or slow rotation \cite{Sotiriou:2013qea}. We only assume stationarity.
\item
Interactions between the scalar and other fields, including the metric, could provide a nontrivial configuration for the scalar field. However, as long as the new interaction terms do not spoil the structure of the perturbation metric  our proof holds. That is, the interaction term only needs to respect shift symmetry upon linearization.
\item Even with the shift-symmetric Horndeski class, there exists a term that inevitably gives rise to scalar hair \cite{Sotiriou:2013qea,Sotiriou:2014pfa}, $\Phi \mathcal{G}$, where $\mathcal{G}=R_{abcd}R^{abcd}-4R_{ab}R^{ab}+R^2$ is the Gauss-Bonnet invariant.
\end{enumerate}
To elaborate on this last point, consider a theory that belongs to the class of theories given by the Lagrangian \eqref{eq:horndeski_lagrangian} and assume $\varphi=const$ is a solution to that theory. Then  add  $\alpha \Phi \mathcal{G}$, where $\alpha$ is a coupling constant, to the Lagrangian. The resulting theory is still in the class of shift-symmetric Horndeski theories, since $\mathcal{G}$ is a total derivative term in 4 dimensions. The new term in the Lagrangian adds $\alpha \mathcal{G}$ to the scalar field's equation of motion and $\mathcal{G}$ does not vanish in black hole spacetimes. Consequently, the new theory must admit a hairy solution only. However,
the new term in the scalar field equation of motion vanishes from the equation of motion for scalar linear perturbations on a {\it fixed} background. Hence, it does not change the structure of the effective metric and our proof applies to this class of theories as well.

Clearly, our central simplifying assumption to consider scalar perturbations on fixed metric backgrounds is a very strong limitation. It is pertinent to revisit the problem taking metric perturbations into account. The perturbative analysis presented in Ref.~\cite{Babichev:2018uiw}, though focused on the role of causality for stability rather than exploring the horizon structure, provides some useful background for extending our analysis.

We also discussed briefly the possibility of having different horizons in configurations where the scalar does not respect stationarity and we provided a simple example in the decoupling limit, in order to motivate further work.  Another crucial piece of motivation to look further into this issue is the following. As has been pointed out in the context of Lorentz-violating theories, where solutions with nested horizons for different excitations are common \cite{Eling:2006ec,Barausse:2011pu,Barausse:2013nwa}, the region between two such horizons can be seen as an ergoregion. In that region, the slowest of the two excitations can carry negative Killing energy as it is behind its Killing horizon and at the same time the faster excitation can still escape. If the two modes interact, then there is a possibility of energy extraction \cite{Dubovsky:2006vk,Eling:2007qd,Jacobson:2008yc}. Assuming that such a process can take place \cite{Benkel:2018abt}, it could lead to thermodynamics conundrums \cite{Dubovsky:2006vk,Eling:2007qd,Jacobson:2008yc}, or perhaps just provide a natural decay channel from hairy to nonhairy solutions \cite{Benkel:2018abt}.

\section{Acknowledgements}\label{sec:acknowledgements}

We are grateful to Norihiro Tanahashi for enlightening discussions. T.P.S. acknowledges partial support from the STFC Consolidated Grant No. ST/P000703/1. M.S. is supported by the Royal Commission for the Exhibition of 1851. M.S. and T.P.S. would like to thank Instituto Superior Tecnico for its hospitality during the final stages of this work. We would also like to acknowledge networking support by the COST Action GWverse CA16104.

\bibliography{notes}

\begin{thebibliography}{41}%
\makeatletter
\providecommand \@ifxundefined [1]{%
 \@ifx{#1\undefined}
}%
\providecommand \@ifnum [1]{%
 \ifnum #1\expandafter \@firstoftwo
 \else \expandafter \@secondoftwo
 \fi
}%
\providecommand \@ifx [1]{%
 \ifx #1\expandafter \@firstoftwo
 \else \expandafter \@secondoftwo
 \fi
}%
\providecommand \natexlab [1]{#1}%
\providecommand \enquote  [1]{``#1''}%
\providecommand \bibnamefont  [1]{#1}%
\providecommand \bibfnamefont [1]{#1}%
\providecommand \citenamefont [1]{#1}%
\providecommand \href@noop [0]{\@secondoftwo}%
\providecommand \href [0]{\begingroup \@sanitize@url \@href}%
\providecommand \@href[1]{\@@startlink{#1}\@@href}%
\providecommand \@@href[1]{\endgroup#1\@@endlink}%
\providecommand \@sanitize@url [0]{\catcode `\\12\catcode `\$12\catcode
  `\&12\catcode `\#12\catcode `\^12\catcode `\_12\catcode `\%12\relax}%
\providecommand \@@startlink[1]{}%
\providecommand \@@endlink[0]{}%
\providecommand \url  [0]{\begingroup\@sanitize@url \@url }%
\providecommand \@url [1]{\endgroup\@href {#1}{\urlprefix }}%
\providecommand \urlprefix  [0]{URL }%
\providecommand \Eprint [0]{\href }%
\providecommand \doibase [0]{http://dx.doi.org/}%
\providecommand \selectlanguage [0]{\@gobble}%
\providecommand \bibinfo  [0]{\@secondoftwo}%
\providecommand \bibfield  [0]{\@secondoftwo}%
\providecommand \translation [1]{[#1]}%
\providecommand \BibitemOpen [0]{}%
\providecommand \bibitemStop [0]{}%
\providecommand \bibitemNoStop [0]{.\EOS\space}%
\providecommand \EOS [0]{\spacefactor3000\relax}%
\providecommand \BibitemShut  [1]{\csname bibitem#1\endcsname}%
\let\auto@bib@innerbib\@empty
\bibitem [{\citenamefont {Israel}(1967)}]{Israel:1967wq}%
  \BibitemOpen
  \bibfield  {author} {\bibinfo {author} {\bibfnamefont {W.}~\bibnamefont
  {Israel}},\ }\href {\doibase 10.1103/PhysRev.164.1776} {\bibfield  {journal}
  {\bibinfo  {journal} {Phys. Rev.}\ }\textbf {\bibinfo {volume} {164}},\
  \bibinfo {pages} {1776} (\bibinfo {year} {1967})}\BibitemShut {NoStop}%
\bibitem [{\citenamefont {Carter}(1971)}]{Carter:1971zc}%
  \BibitemOpen
  \bibfield  {author} {\bibinfo {author} {\bibfnamefont {B.}~\bibnamefont
  {Carter}},\ }\href {\doibase 10.1103/PhysRevLett.26.331} {\bibfield
  {journal} {\bibinfo  {journal} {Phys. Rev. Lett.}\ }\textbf {\bibinfo
  {volume} {26}},\ \bibinfo {pages} {331} (\bibinfo {year} {1971})}\BibitemShut
  {NoStop}%
\bibitem [{\citenamefont {Robinson}(1975)}]{Robinson:1975bv}%
  \BibitemOpen
  \bibfield  {author} {\bibinfo {author} {\bibfnamefont {D.~C.}\ \bibnamefont
  {Robinson}},\ }\href {\doibase 10.1103/PhysRevLett.34.905} {\bibfield
  {journal} {\bibinfo  {journal} {Phys. Rev. Lett.}\ }\textbf {\bibinfo
  {volume} {34}},\ \bibinfo {pages} {905} (\bibinfo {year} {1975})}\BibitemShut
  {NoStop}%
\bibitem [{\citenamefont {Sotiriou}(2015{\natexlab{a}})}]{Sotiriou:2015lxa}%
  \BibitemOpen
  \bibfield  {author} {\bibinfo {author} {\bibfnamefont {T.~P.}\ \bibnamefont
  {Sotiriou}},\ }\bibfield  {booktitle} {\emph {\bibinfo {booktitle}
  {{Proceedings of the 7th Aegean Summer School : Beyond Einstein's theory of
  gravity. Modifications of Einstein's Theory of Gravity at Large Distances.:
  Paros, Greece, September 23-28, 2013}}},\ }\href {\doibase
  10.1007/978-3-319-10070-8_1} {\bibfield  {journal} {\bibinfo  {journal}
  {Lect. Notes Phys.}\ }\textbf {\bibinfo {volume} {892}},\ \bibinfo {pages}
  {3} (\bibinfo {year} {2015}{\natexlab{a}})},\ \Eprint
  {http://arxiv.org/abs/1404.2955} {arXiv:1404.2955 [gr-qc]} \BibitemShut
  {NoStop}%
\bibitem [{\citenamefont {Sotiriou}(2015{\natexlab{b}})}]{Sotiriou:2015pka}%
  \BibitemOpen
  \bibfield  {author} {\bibinfo {author} {\bibfnamefont {T.~P.}\ \bibnamefont
  {Sotiriou}},\ }\href {\doibase 10.1088/0264-9381/32/21/214002} {\bibfield
  {journal} {\bibinfo  {journal} {Class. Quant. Grav.}\ }\textbf {\bibinfo
  {volume} {32}},\ \bibinfo {pages} {214002} (\bibinfo {year}
  {2015}{\natexlab{b}})},\ \Eprint {http://arxiv.org/abs/1505.00248}
  {arXiv:1505.00248 [gr-qc]} \BibitemShut {NoStop}%
\bibitem [{\citenamefont {Horndeski}(1974)}]{Horndeski:1974wa}%
  \BibitemOpen
  \bibfield  {author} {\bibinfo {author} {\bibfnamefont {G.~W.}\ \bibnamefont
  {Horndeski}},\ }\href {\doibase 10.1007/BF01807638} {\bibfield  {journal}
  {\bibinfo  {journal} {Int. J. Theor. Phys.}\ }\textbf {\bibinfo {volume}
  {10}},\ \bibinfo {pages} {363} (\bibinfo {year} {1974})}\BibitemShut
  {NoStop}%
\bibitem [{\citenamefont {Deffayet}\ \emph {et~al.}(2009)\citenamefont
  {Deffayet}, \citenamefont {Deser},\ and\ \citenamefont
  {Esposito-Farèse}}]{Deffayet:2009mn}%
  \BibitemOpen
  \bibfield  {author} {\bibinfo {author} {\bibfnamefont {C.}~\bibnamefont
  {Deffayet}}, \bibinfo {author} {\bibfnamefont {S.}~\bibnamefont {Deser}}, \
  and\ \bibinfo {author} {\bibfnamefont {G.}~\bibnamefont {Esposito-Farèse}},\
  }\href {\doibase 10.1103/PhysRevD.80.064015} {\bibfield  {journal} {\bibinfo
  {journal} {Phys. Rev.}\ }\textbf {\bibinfo {volume} {D80}},\ \bibinfo {pages}
  {064015} (\bibinfo {year} {2009})},\ \Eprint {http://arxiv.org/abs/0906.1967}
  {arXiv:0906.1967 [gr-qc]} \BibitemShut {NoStop}%
\bibitem [{\citenamefont {Kobayashi}\ \emph {et~al.}(2011)\citenamefont
  {Kobayashi}, \citenamefont {Yamaguchi},\ and\ \citenamefont
  {Yokoyama}}]{Kobayashi:2011nu}%
  \BibitemOpen
  \bibfield  {author} {\bibinfo {author} {\bibfnamefont {T.}~\bibnamefont
  {Kobayashi}}, \bibinfo {author} {\bibfnamefont {M.}~\bibnamefont
  {Yamaguchi}}, \ and\ \bibinfo {author} {\bibfnamefont {J.}~\bibnamefont
  {Yokoyama}},\ }\href {\doibase 10.1143/PTP.126.511} {\bibfield  {journal}
  {\bibinfo  {journal} {Prog. Theor. Phys.}\ }\textbf {\bibinfo {volume}
  {126}},\ \bibinfo {pages} {511} (\bibinfo {year} {2011})},\ \Eprint
  {http://arxiv.org/abs/1105.5723} {arXiv:1105.5723 [hep-th]} \BibitemShut
  {NoStop}%
\bibitem [{\citenamefont {{Chase}}(1970)}]{1970CMaPh..19..276C}%
  \BibitemOpen
  \bibfield  {author} {\bibinfo {author} {\bibfnamefont {J.~E.}\ \bibnamefont
  {{Chase}}},\ }\href {\doibase 10.1007/BF01646635} {\bibfield  {journal}
  {\bibinfo  {journal} {Communications in Mathematical Physics}\ }\textbf
  {\bibinfo {volume} {19}},\ \bibinfo {pages} {276} (\bibinfo {year}
  {1970})}\BibitemShut {NoStop}%
\bibitem [{\citenamefont {Hawking}(1972)}]{Hawking:1972qk}%
  \BibitemOpen
  \bibfield  {author} {\bibinfo {author} {\bibfnamefont {S.~W.}\ \bibnamefont
  {Hawking}},\ }\href {\doibase 10.1007/BF01877518} {\bibfield  {journal}
  {\bibinfo  {journal} {Commun. Math. Phys.}\ }\textbf {\bibinfo {volume}
  {25}},\ \bibinfo {pages} {167} (\bibinfo {year} {1972})}\BibitemShut
  {NoStop}%
\bibitem [{\citenamefont {Bekenstein}(1995)}]{Bekenstein:1995un}%
  \BibitemOpen
  \bibfield  {author} {\bibinfo {author} {\bibfnamefont {J.~D.}\ \bibnamefont
  {Bekenstein}},\ }\href {\doibase 10.1103/PhysRevD.51.R6608} {\bibfield
  {journal} {\bibinfo  {journal} {Phys. Rev.}\ }\textbf {\bibinfo {volume}
  {D51}},\ \bibinfo {pages} {R6608} (\bibinfo {year} {1995})}\BibitemShut
  {NoStop}%
\bibitem [{\citenamefont {Sotiriou}\ and\ \citenamefont
  {Faraoni}(2012)}]{Sotiriou:2011dz}%
  \BibitemOpen
  \bibfield  {author} {\bibinfo {author} {\bibfnamefont {T.~P.}\ \bibnamefont
  {Sotiriou}}\ and\ \bibinfo {author} {\bibfnamefont {V.}~\bibnamefont
  {Faraoni}},\ }\href {\doibase 10.1103/PhysRevLett.108.081103} {\bibfield
  {journal} {\bibinfo  {journal} {Phys. Rev. Lett.}\ }\textbf {\bibinfo
  {volume} {108}},\ \bibinfo {pages} {081103} (\bibinfo {year} {2012})},\
  \Eprint {http://arxiv.org/abs/1109.6324} {arXiv:1109.6324 [gr-qc]}
  \BibitemShut {NoStop}%
\bibitem [{\citenamefont {Hui}\ and\ \citenamefont
  {Nicolis}(2013)}]{Hui:2012qt}%
  \BibitemOpen
  \bibfield  {author} {\bibinfo {author} {\bibfnamefont {L.}~\bibnamefont
  {Hui}}\ and\ \bibinfo {author} {\bibfnamefont {A.}~\bibnamefont {Nicolis}},\
  }\href {\doibase 10.1103/PhysRevLett.110.241104} {\bibfield  {journal}
  {\bibinfo  {journal} {Phys. Rev. Lett.}\ }\textbf {\bibinfo {volume} {110}},\
  \bibinfo {pages} {241104} (\bibinfo {year} {2013})},\ \Eprint
  {http://arxiv.org/abs/1202.1296} {arXiv:1202.1296 [hep-th]} \BibitemShut
  {NoStop}%
\bibitem [{\citenamefont {Sotiriou}\ and\ \citenamefont
  {Zhou}(2014{\natexlab{a}})}]{Sotiriou:2013qea}%
  \BibitemOpen
  \bibfield  {author} {\bibinfo {author} {\bibfnamefont {T.~P.}\ \bibnamefont
  {Sotiriou}}\ and\ \bibinfo {author} {\bibfnamefont {S.-Y.}\ \bibnamefont
  {Zhou}},\ }\href {\doibase 10.1103/PhysRevLett.112.251102} {\bibfield
  {journal} {\bibinfo  {journal} {Phys. Rev. Lett.}\ }\textbf {\bibinfo
  {volume} {112}},\ \bibinfo {pages} {251102} (\bibinfo {year}
  {2014}{\natexlab{a}})},\ \Eprint {http://arxiv.org/abs/1312.3622}
  {arXiv:1312.3622 [gr-qc]} \BibitemShut {NoStop}%
\bibitem [{\citenamefont {Sotiriou}\ and\ \citenamefont
  {Zhou}(2014{\natexlab{b}})}]{Sotiriou:2014pfa}%
  \BibitemOpen
  \bibfield  {author} {\bibinfo {author} {\bibfnamefont {T.~P.}\ \bibnamefont
  {Sotiriou}}\ and\ \bibinfo {author} {\bibfnamefont {S.-Y.}\ \bibnamefont
  {Zhou}},\ }\href {\doibase 10.1103/PhysRevD.90.124063} {\bibfield  {journal}
  {\bibinfo  {journal} {Phys. Rev.}\ }\textbf {\bibinfo {volume} {D90}},\
  \bibinfo {pages} {124063} (\bibinfo {year} {2014}{\natexlab{b}})},\ \Eprint
  {http://arxiv.org/abs/1408.1698} {arXiv:1408.1698 [gr-qc]} \BibitemShut
  {NoStop}%
\bibitem [{\citenamefont {Kanti}\ \emph {et~al.}(1996)\citenamefont {Kanti},
  \citenamefont {Mavromatos}, \citenamefont {Rizos}, \citenamefont {Tamvakis},\
  and\ \citenamefont {Winstanley}}]{Kanti:1995vq}%
  \BibitemOpen
  \bibfield  {author} {\bibinfo {author} {\bibfnamefont {P.}~\bibnamefont
  {Kanti}}, \bibinfo {author} {\bibfnamefont {N.~E.}\ \bibnamefont
  {Mavromatos}}, \bibinfo {author} {\bibfnamefont {J.}~\bibnamefont {Rizos}},
  \bibinfo {author} {\bibfnamefont {K.}~\bibnamefont {Tamvakis}}, \ and\
  \bibinfo {author} {\bibfnamefont {E.}~\bibnamefont {Winstanley}},\ }\href
  {\doibase 10.1103/PhysRevD.54.5049} {\bibfield  {journal} {\bibinfo
  {journal} {Phys. Rev.}\ }\textbf {\bibinfo {volume} {D54}},\ \bibinfo {pages}
  {5049} (\bibinfo {year} {1996})},\ \Eprint
  {http://arxiv.org/abs/hep-th/9511071} {arXiv:hep-th/9511071 [hep-th]}
  \BibitemShut {NoStop}%
\bibitem [{\citenamefont {Yunes}\ and\ \citenamefont
  {Stein}(2011)}]{Yunes:2011we}%
  \BibitemOpen
  \bibfield  {author} {\bibinfo {author} {\bibfnamefont {N.}~\bibnamefont
  {Yunes}}\ and\ \bibinfo {author} {\bibfnamefont {L.~C.}\ \bibnamefont
  {Stein}},\ }\href {\doibase 10.1103/PhysRevD.83.104002} {\bibfield  {journal}
  {\bibinfo  {journal} {Phys. Rev.}\ }\textbf {\bibinfo {volume} {D83}},\
  \bibinfo {pages} {104002} (\bibinfo {year} {2011})},\ \Eprint
  {http://arxiv.org/abs/1101.2921} {arXiv:1101.2921 [gr-qc]} \BibitemShut
  {NoStop}%
\bibitem [{\citenamefont {Silva}\ \emph {et~al.}(2018)\citenamefont {Silva},
  \citenamefont {Sakstein}, \citenamefont {Gualtieri}, \citenamefont
  {Sotiriou},\ and\ \citenamefont {Berti}}]{Silva:2017uqg}%
  \BibitemOpen
  \bibfield  {author} {\bibinfo {author} {\bibfnamefont {H.~O.}\ \bibnamefont
  {Silva}}, \bibinfo {author} {\bibfnamefont {J.}~\bibnamefont {Sakstein}},
  \bibinfo {author} {\bibfnamefont {L.}~\bibnamefont {Gualtieri}}, \bibinfo
  {author} {\bibfnamefont {T.~P.}\ \bibnamefont {Sotiriou}}, \ and\ \bibinfo
  {author} {\bibfnamefont {E.}~\bibnamefont {Berti}},\ }\href {\doibase
  10.1103/PhysRevLett.120.131104} {\bibfield  {journal} {\bibinfo  {journal}
  {Phys. Rev. Lett.}\ }\textbf {\bibinfo {volume} {120}},\ \bibinfo {pages}
  {131104} (\bibinfo {year} {2018})},\ \Eprint
  {http://arxiv.org/abs/1711.02080} {arXiv:1711.02080 [gr-qc]} \BibitemShut
  {NoStop}%
\bibitem [{\citenamefont {Doneva}\ and\ \citenamefont
  {Yazadjiev}(2018)}]{Doneva:2017bvd}%
  \BibitemOpen
  \bibfield  {author} {\bibinfo {author} {\bibfnamefont {D.~D.}\ \bibnamefont
  {Doneva}}\ and\ \bibinfo {author} {\bibfnamefont {S.~S.}\ \bibnamefont
  {Yazadjiev}},\ }\href {\doibase 10.1103/PhysRevLett.120.131103} {\bibfield
  {journal} {\bibinfo  {journal} {Phys. Rev. Lett.}\ }\textbf {\bibinfo
  {volume} {120}},\ \bibinfo {pages} {131103} (\bibinfo {year} {2018})},\
  \Eprint {http://arxiv.org/abs/1711.01187} {arXiv:1711.01187 [gr-qc]}
  \BibitemShut {NoStop}%
\bibitem [{\citenamefont {Antoniou}\ \emph {et~al.}(2018)\citenamefont
  {Antoniou}, \citenamefont {Bakopoulos},\ and\ \citenamefont
  {Kanti}}]{Antoniou:2017acq}%
  \BibitemOpen
  \bibfield  {author} {\bibinfo {author} {\bibfnamefont {G.}~\bibnamefont
  {Antoniou}}, \bibinfo {author} {\bibfnamefont {A.}~\bibnamefont
  {Bakopoulos}}, \ and\ \bibinfo {author} {\bibfnamefont {P.}~\bibnamefont
  {Kanti}},\ }\href {\doibase 10.1103/PhysRevLett.120.131102} {\bibfield
  {journal} {\bibinfo  {journal} {Phys. Rev. Lett.}\ }\textbf {\bibinfo
  {volume} {120}},\ \bibinfo {pages} {131102} (\bibinfo {year} {2018})},\
  \Eprint {http://arxiv.org/abs/1711.03390} {arXiv:1711.03390 [hep-th]}
  \BibitemShut {NoStop}%
\bibitem [{\citenamefont {Herdeiro}\ and\ \citenamefont
  {Radu}(2014)}]{Herdeiro:2014goa}%
  \BibitemOpen
  \bibfield  {author} {\bibinfo {author} {\bibfnamefont {C.~A.~R.}\
  \bibnamefont {Herdeiro}}\ and\ \bibinfo {author} {\bibfnamefont
  {E.}~\bibnamefont {Radu}},\ }\href {\doibase 10.1103/PhysRevLett.112.221101}
  {\bibfield  {journal} {\bibinfo  {journal} {Phys. Rev. Lett.}\ }\textbf
  {\bibinfo {volume} {112}},\ \bibinfo {pages} {221101} (\bibinfo {year}
  {2014})},\ \Eprint {http://arxiv.org/abs/1403.2757} {arXiv:1403.2757 [gr-qc]}
  \BibitemShut {NoStop}%
\bibitem [{\citenamefont {Babichev}\ and\ \citenamefont
  {Charmousis}(2014)}]{Babichev:2013cya}%
  \BibitemOpen
  \bibfield  {author} {\bibinfo {author} {\bibfnamefont {E.}~\bibnamefont
  {Babichev}}\ and\ \bibinfo {author} {\bibfnamefont {C.}~\bibnamefont
  {Charmousis}},\ }\href {\doibase 10.1007/JHEP08(2014)106} {\bibfield
  {journal} {\bibinfo  {journal} {JHEP}\ }\textbf {\bibinfo {volume} {08}},\
  \bibinfo {pages} {106} (\bibinfo {year} {2014})},\ \Eprint
  {http://arxiv.org/abs/1312.3204} {arXiv:1312.3204 [gr-qc]} \BibitemShut
  {NoStop}%
\bibitem [{\citenamefont {Charmousis}\ \emph {et~al.}(2014)\citenamefont
  {Charmousis}, \citenamefont {Kolyvaris}, \citenamefont {Papantonopoulos},\
  and\ \citenamefont {Tsoukalas}}]{Charmousis:2014zaa}%
  \BibitemOpen
  \bibfield  {author} {\bibinfo {author} {\bibfnamefont {C.}~\bibnamefont
  {Charmousis}}, \bibinfo {author} {\bibfnamefont {T.}~\bibnamefont
  {Kolyvaris}}, \bibinfo {author} {\bibfnamefont {E.}~\bibnamefont
  {Papantonopoulos}}, \ and\ \bibinfo {author} {\bibfnamefont {M.}~\bibnamefont
  {Tsoukalas}},\ }\href {\doibase 10.1007/JHEP07(2014)085} {\bibfield
  {journal} {\bibinfo  {journal} {JHEP}\ }\textbf {\bibinfo {volume} {07}},\
  \bibinfo {pages} {085} (\bibinfo {year} {2014})},\ \Eprint
  {http://arxiv.org/abs/1404.1024} {arXiv:1404.1024 [gr-qc]} \BibitemShut
  {NoStop}%
\bibitem [{\citenamefont {Kobayashi}\ and\ \citenamefont
  {Tanahashi}(2014)}]{Kobayashi:2014eva}%
  \BibitemOpen
  \bibfield  {author} {\bibinfo {author} {\bibfnamefont {T.}~\bibnamefont
  {Kobayashi}}\ and\ \bibinfo {author} {\bibfnamefont {N.}~\bibnamefont
  {Tanahashi}},\ }\href {\doibase 10.1093/ptep/ptu096} {\bibfield  {journal}
  {\bibinfo  {journal} {PTEP}\ }\textbf {\bibinfo {volume} {2014}},\ \bibinfo
  {pages} {073E02} (\bibinfo {year} {2014})},\ \Eprint
  {http://arxiv.org/abs/1403.4364} {arXiv:1403.4364 [gr-qc]} \BibitemShut
  {NoStop}%
\bibitem [{\citenamefont {Babichev}\ \emph {et~al.}(2008)\citenamefont
  {Babichev}, \citenamefont {Mukhanov},\ and\ \citenamefont
  {Vikman}}]{Babichev:2007dw}%
  \BibitemOpen
  \bibfield  {author} {\bibinfo {author} {\bibfnamefont {E.}~\bibnamefont
  {Babichev}}, \bibinfo {author} {\bibfnamefont {V.}~\bibnamefont {Mukhanov}},
  \ and\ \bibinfo {author} {\bibfnamefont {A.}~\bibnamefont {Vikman}},\ }\href
  {\doibase 10.1088/1126-6708/2008/02/101} {\bibfield  {journal} {\bibinfo
  {journal} {JHEP}\ }\textbf {\bibinfo {volume} {02}},\ \bibinfo {pages} {101}
  (\bibinfo {year} {2008})},\ \Eprint {http://arxiv.org/abs/0708.0561}
  {arXiv:0708.0561 [hep-th]} \BibitemShut {NoStop}%
\bibitem [{\citenamefont {Babichev}\ \emph {et~al.}(2017)\citenamefont
  {Babichev}, \citenamefont {Charmousis}, \citenamefont {Esposito-Farèse},\
  and\ \citenamefont {Lehébel}}]{Babichev:2017lmw}%
  \BibitemOpen
  \bibfield  {author} {\bibinfo {author} {\bibfnamefont {E.}~\bibnamefont
  {Babichev}}, \bibinfo {author} {\bibfnamefont {C.}~\bibnamefont
  {Charmousis}}, \bibinfo {author} {\bibfnamefont {G.}~\bibnamefont
  {Esposito-Farèse}}, \ and\ \bibinfo {author} {\bibfnamefont
  {A.}~\bibnamefont {Lehébel}},\ }\href@noop {} {\  (\bibinfo {year}
  {2017})},\ \Eprint {http://arxiv.org/abs/1712.04398} {arXiv:1712.04398
  [gr-qc]} \BibitemShut {NoStop}%
\bibitem [{\citenamefont {Babichev}\ \emph {et~al.}(2018)\citenamefont
  {Babichev}, \citenamefont {Charmousis}, \citenamefont {Esposito-Farèse},\
  and\ \citenamefont {Lehébel}}]{Babichev:2018uiw}%
  \BibitemOpen
  \bibfield  {author} {\bibinfo {author} {\bibfnamefont {E.}~\bibnamefont
  {Babichev}}, \bibinfo {author} {\bibfnamefont {C.}~\bibnamefont
  {Charmousis}}, \bibinfo {author} {\bibfnamefont {G.}~\bibnamefont
  {Esposito-Farèse}}, \ and\ \bibinfo {author} {\bibfnamefont
  {A.}~\bibnamefont {Lehébel}},\ }\href@noop {} {\  (\bibinfo {year}
  {2018})},\ \Eprint {http://arxiv.org/abs/1803.11444} {arXiv:1803.11444
  [gr-qc]} \BibitemShut {NoStop}%
\bibitem [{\citenamefont {Tanahashi}\ and\ \citenamefont
  {Ohashi}(2017)}]{Tanahashi:2017kgn}%
  \BibitemOpen
  \bibfield  {author} {\bibinfo {author} {\bibfnamefont {N.}~\bibnamefont
  {Tanahashi}}\ and\ \bibinfo {author} {\bibfnamefont {S.}~\bibnamefont
  {Ohashi}},\ }\href {\doibase 10.1088/1361-6382/aa85fb} {\bibfield  {journal}
  {\bibinfo  {journal} {Class. Quant. Grav.}\ }\textbf {\bibinfo {volume}
  {34}},\ \bibinfo {pages} {215003} (\bibinfo {year} {2017})},\ \Eprint
  {http://arxiv.org/abs/1704.02757} {arXiv:1704.02757 [hep-th]} \BibitemShut
  {NoStop}%
\bibitem [{\citenamefont {Minamitsuji}(2015)}]{Minamitsuji:2015nca}%
  \BibitemOpen
  \bibfield  {author} {\bibinfo {author} {\bibfnamefont {M.}~\bibnamefont
  {Minamitsuji}},\ }\href {\doibase 10.1016/j.physletb.2015.02.064} {\bibfield
  {journal} {\bibinfo  {journal} {Phys. Lett.}\ }\textbf {\bibinfo {volume}
  {B743}},\ \bibinfo {pages} {272} (\bibinfo {year} {2015})}\BibitemShut
  {NoStop}%
\bibitem [{\citenamefont {Reall}\ \emph {et~al.}(2014)\citenamefont {Reall},
  \citenamefont {Tanahashi},\ and\ \citenamefont {Way}}]{Reall:2014pwa}%
  \BibitemOpen
  \bibfield  {author} {\bibinfo {author} {\bibfnamefont {H.}~\bibnamefont
  {Reall}}, \bibinfo {author} {\bibfnamefont {N.}~\bibnamefont {Tanahashi}}, \
  and\ \bibinfo {author} {\bibfnamefont {B.}~\bibnamefont {Way}},\ }\href
  {\doibase 10.1088/0264-9381/31/20/205005} {\bibfield  {journal} {\bibinfo
  {journal} {Class. Quant. Grav.}\ }\textbf {\bibinfo {volume} {31}},\ \bibinfo
  {pages} {205005} (\bibinfo {year} {2014})},\ \Eprint
  {http://arxiv.org/abs/1406.3379} {arXiv:1406.3379 [hep-th]} \BibitemShut
  {NoStop}%
\bibitem [{\citenamefont {Izumi}(2014)}]{Izumi:2014loa}%
  \BibitemOpen
  \bibfield  {author} {\bibinfo {author} {\bibfnamefont {K.}~\bibnamefont
  {Izumi}},\ }\href {\doibase 10.1103/PhysRevD.90.044037} {\bibfield  {journal}
  {\bibinfo  {journal} {Phys. Rev.}\ }\textbf {\bibinfo {volume} {D90}},\
  \bibinfo {pages} {044037} (\bibinfo {year} {2014})},\ \Eprint
  {http://arxiv.org/abs/1406.0677} {arXiv:1406.0677 [gr-qc]} \BibitemShut
  {NoStop}%
\bibitem [{\citenamefont {Carter}(1973)}]{Carter:1973rla}%
  \BibitemOpen
  \bibfield  {author} {\bibinfo {author} {\bibfnamefont {B.}~\bibnamefont
  {Carter}},\ }in\ \href@noop {} {\emph {\bibinfo {booktitle} {{Proceedings,
  Ecole d'Eté de Physique Théorique: Les Astres Occlus: Les Houches, France,
  August, 1972}}}}\ (\bibinfo {year} {1973})\ pp.\ \bibinfo {pages}
  {57--214}\BibitemShut {NoStop}%
\bibitem [{\citenamefont {Ogawa}\ \emph {et~al.}(2016)\citenamefont {Ogawa},
  \citenamefont {Kobayashi},\ and\ \citenamefont {Suyama}}]{Ogawa:2015pea}%
  \BibitemOpen
  \bibfield  {author} {\bibinfo {author} {\bibfnamefont {H.}~\bibnamefont
  {Ogawa}}, \bibinfo {author} {\bibfnamefont {T.}~\bibnamefont {Kobayashi}}, \
  and\ \bibinfo {author} {\bibfnamefont {T.}~\bibnamefont {Suyama}},\ }\href
  {\doibase 10.1103/PhysRevD.93.064078} {\bibfield  {journal} {\bibinfo
  {journal} {Phys. Rev.}\ }\textbf {\bibinfo {volume} {D93}},\ \bibinfo {pages}
  {064078} (\bibinfo {year} {2016})},\ \Eprint
  {http://arxiv.org/abs/1510.07400} {arXiv:1510.07400 [gr-qc]} \BibitemShut
  {NoStop}%
\bibitem [{\citenamefont {Eling}\ and\ \citenamefont
  {Jacobson}(2006)}]{Eling:2006ec}%
  \BibitemOpen
  \bibfield  {author} {\bibinfo {author} {\bibfnamefont {C.}~\bibnamefont
  {Eling}}\ and\ \bibinfo {author} {\bibfnamefont {T.}~\bibnamefont
  {Jacobson}},\ }\href {\doibase 10.1088/0264-9381/23/18/009,
  10.1088/0264-9381/27/4/049802} {\bibfield  {journal} {\bibinfo  {journal}
  {Class. Quant. Grav.}\ }\textbf {\bibinfo {volume} {23}},\ \bibinfo {pages}
  {5643} (\bibinfo {year} {2006})},\ \bibinfo {note} {[Erratum: Class. Quant.
  Grav. {\bf 27}, 049802 (2010)]},\ \Eprint
  {http://arxiv.org/abs/gr-qc/0604088} {arXiv:gr-qc/0604088 [gr-qc]}
  \BibitemShut {NoStop}%
\bibitem [{\citenamefont {Barausse}\ \emph {et~al.}(2011)\citenamefont
  {Barausse}, \citenamefont {Jacobson},\ and\ \citenamefont
  {Sotiriou}}]{Barausse:2011pu}%
  \BibitemOpen
  \bibfield  {author} {\bibinfo {author} {\bibfnamefont {E.}~\bibnamefont
  {Barausse}}, \bibinfo {author} {\bibfnamefont {T.}~\bibnamefont {Jacobson}},
  \ and\ \bibinfo {author} {\bibfnamefont {T.~P.}\ \bibnamefont {Sotiriou}},\
  }\href {\doibase 10.1103/PhysRevD.83.124043} {\bibfield  {journal} {\bibinfo
  {journal} {Phys. Rev. D}\ }\textbf {\bibinfo {volume} {83}},\ \bibinfo
  {pages} {124043} (\bibinfo {year} {2011})},\ \Eprint
  {http://arxiv.org/abs/1104.2889} {arXiv:1104.2889 [gr-qc]} \BibitemShut
  {NoStop}%
\bibitem [{\citenamefont {Barausse}\ and\ \citenamefont
  {Sotiriou}(2013)}]{Barausse:2013nwa}%
  \BibitemOpen
  \bibfield  {author} {\bibinfo {author} {\bibfnamefont {E.}~\bibnamefont
  {Barausse}}\ and\ \bibinfo {author} {\bibfnamefont {T.~P.}\ \bibnamefont
  {Sotiriou}},\ }\href {\doibase 10.1088/0264-9381/30/24/244010} {\bibfield
  {journal} {\bibinfo  {journal} {Class. Quant. Grav.}\ }\textbf {\bibinfo
  {volume} {30}},\ \bibinfo {pages} {244010} (\bibinfo {year} {2013})},\
  \Eprint {http://arxiv.org/abs/1307.3359} {arXiv:1307.3359 [gr-qc]}
  \BibitemShut {NoStop}%
\bibitem [{\citenamefont {Dubovsky}\ and\ \citenamefont
  {Sibiryakov}(2006)}]{Dubovsky:2006vk}%
  \BibitemOpen
  \bibfield  {author} {\bibinfo {author} {\bibfnamefont {S.~L.}\ \bibnamefont
  {Dubovsky}}\ and\ \bibinfo {author} {\bibfnamefont {S.~M.}\ \bibnamefont
  {Sibiryakov}},\ }\href {\doibase 10.1016/j.physletb.2006.05.074} {\bibfield
  {journal} {\bibinfo  {journal} {Phys. Lett.}\ }\textbf {\bibinfo {volume}
  {B638}},\ \bibinfo {pages} {509} (\bibinfo {year} {2006})},\ \Eprint
  {http://arxiv.org/abs/hep-th/0603158} {arXiv:hep-th/0603158 [hep-th]}
  \BibitemShut {NoStop}%
\bibitem [{\citenamefont {Eling}\ \emph {et~al.}(2007)\citenamefont {Eling},
  \citenamefont {Foster}, \citenamefont {Jacobson},\ and\ \citenamefont
  {Wall}}]{Eling:2007qd}%
  \BibitemOpen
  \bibfield  {author} {\bibinfo {author} {\bibfnamefont {C.}~\bibnamefont
  {Eling}}, \bibinfo {author} {\bibfnamefont {B.~Z.}\ \bibnamefont {Foster}},
  \bibinfo {author} {\bibfnamefont {T.}~\bibnamefont {Jacobson}}, \ and\
  \bibinfo {author} {\bibfnamefont {A.~C.}\ \bibnamefont {Wall}},\ }\href
  {\doibase 10.1103/PhysRevD.75.101502} {\bibfield  {journal} {\bibinfo
  {journal} {Phys. Rev. D}\ }\textbf {\bibinfo {volume} {75}},\ \bibinfo
  {pages} {101502} (\bibinfo {year} {2007})},\ \Eprint
  {http://arxiv.org/abs/hep-th/0702124} {arXiv:hep-th/0702124 [hep-th]}
  \BibitemShut {NoStop}%
\bibitem [{\citenamefont {Jacobson}\ and\ \citenamefont
  {Wall}(2010)}]{Jacobson:2008yc}%
  \BibitemOpen
  \bibfield  {author} {\bibinfo {author} {\bibfnamefont {T.}~\bibnamefont
  {Jacobson}}\ and\ \bibinfo {author} {\bibfnamefont {A.~C.}\ \bibnamefont
  {Wall}},\ }\href {\doibase 10.1007/s10701-010-9423-5} {\bibfield  {journal}
  {\bibinfo  {journal} {Found. Phys.}\ }\textbf {\bibinfo {volume} {40}},\
  \bibinfo {pages} {1076} (\bibinfo {year} {2010})},\ \Eprint
  {http://arxiv.org/abs/0804.2720} {arXiv:0804.2720 [hep-th]} \BibitemShut
  {NoStop}%
\bibitem [{\citenamefont {Benkel}\ \emph {et~al.}(2018)\citenamefont {Benkel},
  \citenamefont {Bhattacharyya}, \citenamefont {Louko}, \citenamefont
  {Mattingly},\ and\ \citenamefont {Sotiriou}}]{Benkel:2018abt}%
  \BibitemOpen
  \bibfield  {author} {\bibinfo {author} {\bibfnamefont {R.}~\bibnamefont
  {Benkel}}, \bibinfo {author} {\bibfnamefont {J.}~\bibnamefont
  {Bhattacharyya}}, \bibinfo {author} {\bibfnamefont {J.}~\bibnamefont
  {Louko}}, \bibinfo {author} {\bibfnamefont {D.}~\bibnamefont {Mattingly}}, \
  and\ \bibinfo {author} {\bibfnamefont {T.~P.}\ \bibnamefont {Sotiriou}},\
  }\href@noop {} {\  (\bibinfo {year} {2018})},\ \Eprint
  {http://arxiv.org/abs/1803.01624} {arXiv:1803.01624 [gr-qc]} \BibitemShut
  {NoStop}%
\bibitem [{\citenamefont {Gourgoulhon}\ and\ \citenamefont
  {Jaramillo}(2006)}]{Gourgoulhon:2005ng}%
  \BibitemOpen
  \bibfield  {author} {\bibinfo {author} {\bibfnamefont {E.}~\bibnamefont
  {Gourgoulhon}}\ and\ \bibinfo {author} {\bibfnamefont {J.~L.}\ \bibnamefont
  {Jaramillo}},\ }\href {\doibase 10.1016/j.physrep.2005.10.005} {\bibfield
  {journal} {\bibinfo  {journal} {Phys. Rept.}\ }\textbf {\bibinfo {volume}
  {423}},\ \bibinfo {pages} {159} (\bibinfo {year} {2006})},\ \Eprint
  {http://arxiv.org/abs/gr-qc/0503113} {arXiv:gr-qc/0503113 [gr-qc]}
  \BibitemShut {NoStop}%
\end{thebibliography}%

\appendix

\appendix

\section{Effective metric in shift-symmetric Horndeski}\label{appendix_effective_metric}
In this section we show the exact form of the effective metric defined in Eq.~\eqref{eq:effective metric definition}. We split
\begin{equation}
f^{ab}=\sum_{n=2}^{5}f_{(n)}^{ab},
\end{equation}
where the subscript index $n$ refers to the contribution coming from $\L_n$, and use the simplifying notation $\nabla_a \varphi = \varphi_a$ and $\nabla_b \nabla_a \varphi = \varphi_{ab}$. We have explicitly
\onecolumngrid

\begin{align}
  f_{(2)}^{ab} &= G_{2 X} g^{ab} - G_{2 X X} \varphi^{a} \varphi^{b}\,, \\
  f_{(3)}^{ab} &= 2 G_{3X} \left( \varphi^{ab} - \DAlembert \varphi g^{ab}\right)+G_{3XX} \left(\DAlembert \varphi \varphi^{a} \varphi^{b} + \varphi_{c} \varphi_{d} \varphi^{cd} g^{ab} - \varphi^{ca} \varphi^{b} \varphi_{c}- \varphi^{cb} \varphi^{a} \varphi_{c}\right)\,, \\
  f_{(4)}^{ab} &= G_{4X} f^{ab}_{(4,1)} + G_{4XX} f^{ab}_{(4,2)} + G_{4XXX} f^{ab}_{(4,3)}\,, \\
  f_{(5)}^{ab} &= G_{5X} f^{ab}_{(5,1)} + G_{5XX} f^{ab}_{(5,2)} + G_{5XXX} f^{ab}_{(5,3)}\,,
\end{align}
where
\begin{align}
  f_{(4,1)}^{ab} =& - 2 G^{ab}\,, \\
  f_{(4,2)}^{ab} =& - R \varphi^{a} \varphi^{b} + 3 (\DAlembert \varphi)^{2} g^{ab} - 3 (\varphi_{cd})^{2} g^{ab} - 6 \DAlembert \varphi \varphi^{ab} - 2 R_{cd} \varphi^{c} \varphi^{d} g^{ab} + 2 R^{acbd} \varphi_{c} \varphi_{d} + 4 R^{ac} \varphi_{c} \varphi^{b} + 6 \varphi^{ac} {\varphi^{b}}_c\,, \\
  f_{(4,3)}^{ab} =& - 2 \DAlembert \varphi \varphi_{cd} \varphi^{c} \varphi^{d} g^{ab} - (\DAlembert \varphi)^{2} \varphi^{a} \varphi^{b} + 2 \varphi_{cd} \varphi^{c} \varphi^{d} \varphi^{ab} + (\varphi_{cd})^{2} \varphi^{a} \varphi^{b}+
			\\   \nonumber
			      & + 2 \varphi^{cd} \varphi_{ed} \varphi_{c} \varphi^{e} g^{ab}+ 4 \DAlembert \varphi \varphi^{ac} \varphi_{c} \varphi^{b} - 2 \varphi^{ac} \varphi^{bd} \varphi_{c} \varphi_{d} - 4 \varphi_{cd} \varphi^{b} \varphi^{d} \varphi^{ac} \,, \\
  f_{(5,1)}^{ab} =&\, \DAlembert \varphi G^{ab} + G_{cd} \varphi^{cd} g^{ab} - 2 G^{ac} {\varphi_{c}}^{b} + \DAlembert \varphi R^{ab} + R_{cd} \varphi^{cd} g^{ab} - 2 {\varphi^{a}}_{c} R^{bc} + 2 \varphi_{cd} R^{acdb}\,, \\
  f_{(5,2)}^{ab} =& - G^{ab} \varphi_{c} \varphi_{d} \varphi^{cd} - G_{cd} \varphi^{cd} \varphi^{a} \varphi^{b}
			+ 2 G^{ac} \varphi^{b} \varphi_{cd} \varphi^{d}
			+ \DAlembert \varphi \left( (\varphi_{cd})^{2} + \varphi^{c} \varphi^{d} R_{cd} \right) g^{ab}
			+ (\DAlembert \varphi)^{2} \varphi^{ab}+
			\\ \nonumber
			&- 2 \varphi_{cd} \varphi^{d} R^{ce} \varphi_{e} g^{ab}
			- 2 \DAlembert \varphi R^{ac}  \varphi_{c} \varphi^{b}
			- \varphi^{ab} \left( 2 (\varphi_{cd})^{2} + \varphi^{c} \varphi^{d} R_{cd} \right)
			- 2 \varphi_{cd} \varphi^{d} \varphi_{e} R^{beca}
			- 2 \varphi_{cd} \varphi^{b} \varphi_{e} R^{deac}+
			\\ \nonumber
			&- g^{ab} \varphi^{cd} \left( \varphi_{ce} {\varphi^{e}}_{d} - \varphi^{e} \varphi^{f} R_{cefd} \right)
			- \DAlembert \varphi ( 3 \varphi^{ac} {\varphi^{b}}_{c} - \varphi_{c} \varphi_{d} R^{acdb} )
			+ 2 \varphi^{ac} \varphi^{b} R_{cd} \varphi^{d}
			+ 2 R^{ac} \varphi_{c} \varphi^{bd} \varphi_{d}+
			\\ \nonumber
			&+ 2 \varphi^{cb} (2 \varphi^{ad} \varphi_{cd} + \varphi^{d} \varphi_{e} {R^{ae}}_{cd} )
			- \frac{1}{6} g^{ab} \left( 4 (\DAlembert \varphi)^{3} - 6 \DAlembert \varphi (\varphi_{cd})^{2} + 2 (\varphi_{cd})^{3} \right)+
			\\ \nonumber
			&+ (\DAlembert \varphi)^{2} \varphi^{ab}
			- \DAlembert \varphi \varphi^{ac} {\varphi^{b}}_{c}\,,  \\
  f_{(5,3)}^{ab} =& - (\DAlembert \varphi)^{2} \varphi^{ac} \varphi_{c} \varphi^{b} - \DAlembert \varphi \varphi_{cd} \varphi^{d} \varphi^{ce} \varphi_{e} g^{ab}
			+ \varphi_{cd} \varphi^{c} \varphi^{de} \varphi_{e} \varphi^{ab} + (\varphi_{de})^{2} \varphi^{ac} \varphi_{c} \varphi^{b}
			+ \varphi^{cd} \varphi_{ce} \varphi^{e} \varphi_{df} \varphi^{f} g^{ab}+
			\\ \nonumber
			&+ \DAlembert \varphi \varphi^{ac} \varphi_{c} \varphi^{bd} \varphi_{d}
			+ 2 \DAlembert \varphi \varphi^{ac} \varphi^{b} \varphi_{cd} \varphi^{d}
			- 2 \varphi^{ad} \varphi_{d} \varphi^{bc} \varphi_{ce} \varphi^{e}
			- 2 \varphi^{ad} \varphi_{dc} \varphi^{b} \varphi^{ce} \varphi_{e}+
			\\ \nonumber
			&+ \frac{1}{6} \left( (\DAlembert \varphi)^{3} - 3 \DAlembert \varphi (\varphi_{cd})^{2} + 2 (\varphi_{cd})^{3} \right) \varphi^{a} \varphi^{b}
			+ \frac{1}{2} (\DAlembert \varphi)^{2} \varphi_{c} \varphi_{d} \varphi^{cd} g^{ab}+
			\\ \nonumber
			&- \frac{1}{2} \varphi_{c} \varphi_{d} \varphi^{cd} (\varphi_{ef})^{2} g^{ab}
			- \DAlembert \varphi \varphi_{c} \varphi_{d} \varphi^{cd} \varphi^{ab}
			+ \varphi_{c} \varphi_{d} \varphi^{cd} \varphi^{ae} {\varphi^{b}}_{e}\,.
\end{align}

\twocolumngrid

\section{Killing horizon of $f^{-1}_{ab}$}\label{identity_proof}
Here, we prove Eqs. \eqref{cond1} and \eqref{cond2}. For the proof of Eq. \eqref{cond1} we only use stationarity of the metric and the scalar field. In the proof of Eq. \eqref{cond2}, we also assume that the Killing horizon has constant surface gravity. We use the notation introduced in Appendix \ref{appendix_effective_metric} for simplicity, namely $\varphi_a = \nabla_a \varphi$ and $\varphi_{ab} = \nabla_b \nabla_a \varphi$.

\subsection{Proof of condition I}
The proof of condition I, $f^{ab} \xi_a \xi_b \eqh 0$, is rather straightforward, but it involves going through each term of $f^{ab}$ in Appendix \ref{appendix_effective_metric} to show it vanishes. For an interested reader, the following identities are helpful to reproduce the result:
\bea
g^{ab}\xi_a \xi_b \eqh 0\notag\\
\xi^a \varphi_a = 0\notag\\
\varphi^{ab}\xi_a \proph \xi^b\notag\\
R^{ab} \xi_a \xi_b \eqh 0\notag\\
R_{acbd}\xi^a \eta^c \xi^b \hat \eta^d \eqh 0,\notag
\eea
where $\eta$ and $\hat \eta$ are tangential spacelike directions on the horizon.

\subsection{Proof of condition II}
Let us introduce the following definition:

\beq
p^a \equivh q^a \iff p^a \eqh q^a + c~ \xi^a,
\eeq
i.e. two (co)vectors are equivalent if they differ by a multiple of $\xi^{a}$.

Then, we can express Eq. \eqref{cond2} as
\beq\label{app:mastereq2}
f^{ab}\xi_a \equivh 0.
\eeq
Given the expression for $f^{ab}$ (Appendix \ref{appendix_effective_metric}), the  following three relations
\bea
\xi_a \nabla^b \nabla^a \varphi \equivh 0,\label{app:subeq1}\\
\xi_a R^{ab} \equivh 0,\label{app:subeq2}\\
 \xi_a R^{cbda} T_{cd}\equivh 0\label{app:subeq3},
\eea
where $T_{cd}$ is any symmetric tensor satisfying $T_{cd}\xi^d \equivh 0$, are sufficient for Eq. \eqref{app:mastereq2} to be satisfied.

In what follows, we prove the relations in Eqs. (\ref{app:subeq1}-\ref{app:subeq3}). In the proof of \eqref{app:subeq2} and \eqref{app:subeq3} we assume that the surface gravity of the horizon $\H$ is constant.

\subsubsection{$\xi_a \varphi^{ab} \equivh 0$.}
The scalar field satisfies stationarity, i.e. $\xi^a \varphi_a = 0$.
Taking a derivative of this condition, we get
\beq\label{lemma1:eq1}
\xi_a \varphi^{ab} = -\varphi_a \nabla^b \xi^{a}.
\eeq
We can use the expansion of $\nabla^b\xi^{a}$ on the horizon
\beq
\nabla^b\xi^{a} \eqh -\kappa (\xi^a k^b -\xi^b k^a)+\xi^a \eta^b - \xi^b \eta^a,
\eeq
where $k^a$ is a null direction transverse to $\H$ satisfying $\xi^a k_a =-1$, $\eta^a$ is a spacelike direction orthogonal to $\xi^a$ and $k^a$ and $\kappa$ is the surface gravity of $\H$. Equation \eqref{lemma1:eq1} yields
\beq
\xi_a \varphi^{ab} \eqh -\kappa k_a \varphi^a \xi^b + \eta_a \varphi^a \xi^b \equivh 0,
\eeq

\subsubsection{$\xi_a R^{ab} \equivh 0$.}

The Killing horizon $\H$ has zero expansion, shear and vorticity.  Thus applying Raychaudhuri equation on $\H$ yields
\beq\label{lemma2:eq1}
R_{ab} \xi^a \xi^b \eqh 0.
\eeq
Moreover, for any tangent spacelike direction $\hat \eta^a$ on $\H$
\beq
\kappa_{,a}\hat \eta^a \eqh -R_{ab} \xi^a \hat \eta^b.
\eeq
Thus for a constant surface gravity of $\H$
\beq\label{lemma2:eq2}
R_{ab}\xi^a \hat \eta^b \eqh 0.
\eeq
Eq. \eqref{lemma2:eq1} implies that $R^{ab}\xi_a$ is tangent to $\H$ and eq. \eqref{lemma2:eq2} implies that this vector is orthogonal to all spacelike directions on $\H$. The only vector on $\H$ satisfying these properties has to be a multiple of $\xi^b$. Consequently,
\beq
R^{ab}\xi_a \equivh 0,
\eeq

\subsubsection{$\xi_a R^{cbda} T_{cd}\equivh 0$.}
In order to prove Eq. \eqref{app:subeq3}, we consider a family of null hypersurfaces in an open neighborhood around $\H$ with tangent null direction $l^a$ such that
\beq
l^a \eqh \xi^a.
\eeq
As a result,
\beq\label{lemma3:eq1}
T^{cd}R_{cbda}\xi^a \eqh T^{cd}R_{cbda}l^a = T^{cd}\left(\nabla_c \nabla_b l_{d}-\nabla_b \nabla_c l_{d}\right).
\eeq
By the properties of null hypersurfaces,
\beq\label{lemma3:eq2}
\nabla_b l_{d}= \theta_{bd}+\omega_b l_d -l_b k^c \nabla_c l_{d}
\eeq
where $k^a$ is the null direction transverse to the null hypersurfaces satisfying $l.k = -1$, $\theta_{ab}$ is the projection of $\nabla_b l_{a}$ onto 2-dimensional spacelike submanifold of the null hypersurfaces (orthogonal to $l^a$ and $k^a$) and $\omega_a = l^b \nabla_b k_{a}$ (Eq. (5.20) of \cite{Gourgoulhon:2005ng}).

Taking the derivative of Eq. \eqref{lemma3:eq2} and substituting in Eq. \eqref{lemma3:eq1}, we arrive at
\bea
T^{cd}R_{cbda}\xi^a && \eqh \,T^{cd}\left( \nabla_c \theta_{bd} - \nabla_b \theta_{cd}\right)\notag\\
&&+\,T^{cd}\left(\nabla_c \omega_{b}-\nabla_b\omega_{c} \right)l_d\notag\\
&&+\,T^{cd}\left( \omega_b \nabla_c l_{d}- \omega_c \nabla_b l_{d}\right)\notag\\
&&-\,T^{cd}\left(\nabla_c l_{b}-\nabla_b l_{c} \right)k^e \nabla_e l_{d}\notag\\
&&-\,T^{cd}\left(l_b \nabla_c -l_c \nabla_b \right)(k^e \nabla_e l_{d}).\label{lemma3:eq3}
\eea
We show that each line in the above equation gives a contribution proportional to $\xi_b$ on $\H$.
\begin{enumerate}[(i)]
\item Let us define $u^{(1)}_b = T^{cd}\left( \nabla_c \theta_{bd} - \nabla_b \theta_{cd}\right)$.
We prove $u^{(1)}_b \equivh 0$ by showing $u^{(1)}_b l^b \eqh 0$ and $u^{(1)}_b \eta^b \eqh 0$ where $ \eta^{b}$ is any tangent spacelike direction on $\H$.

Note that $l^b \theta_{ab} =0$ and $\theta_{ab} \eqh 0$. Thus
\bea
l^b \nabla_b \theta_{cd} &&\eqh 0,\\
l^b \nabla_c \theta_{bd} &&= - \left(\nabla_c l^b\right) \theta_{bd} \eqh 0,
\eea
which yields $u^{(1)}_b l^b \eqh 0$.

In order to prove $u^{(1)}_b \eta^b \eqh 0$, notice that
\beq
\eta^b \nabla_b \theta_{cd}\eqh 0.
\eeq
As a result,
\beq
u^{(1)}_b \eta^b \eqh T^{cd}\eta^b \nabla_c \theta_{bd}.
\eeq
From $T^{cd}l_c \equivh 0$, it follows
\beq\label{app:Texpansion}
T^{cd}l_c \eqh \mathcal{T}\,l^d,
\eeq
where $\mathcal{T}$ is a scalar. Thus, we can express $T^{cd}$ as
\beq
T^{cd} \eqh \hat T^{cd} -\mathcal{T}\, k^c l^d.
\eeq
where $\hat T^{cd}l_c \eqh 0$. Consequently,
\beq
T^{cd} \nabla_c \theta_{bd} \eqh \hat T^{cd} \nabla_c \theta_{bd}-\mathcal{T}\, k^c l^d \nabla_c \theta_{bd}
\eeq
The index $c$ of $\hat T^{cd}$ has to be along a tangent direction to $\H$ which implies $\hat T^{cd}\nabla_c \theta_{bd} \eqh 0$. Moreover, $l^d \nabla_c \theta_{bd}\eqh 0$ and this proves $u^{(1)}_b \eta^b \eqh 0$.

\item If we define $u^{(2)}_b = T^{cd}\left(\nabla_c \omega_{b}-\nabla_b \omega_{c;b} \right)l_d$, by Eq. \eqref{app:Texpansion}
\beq
u^{(2)}_b \eqh \mathcal{T} l^c(\nabla_c\omega_{b}-\nabla_b \omega_{c}).
\eeq
Again, one can show
\bea
u^{(2)}_b l^b &&\eqh \mathcal{T} l^b l^c (\nabla_c \omega_{b}-\nabla_b \omega_{c}) =0,\label{app:line2_eq1}\\
u^{(2)}_b \eta^b &&\eqh \mathcal{T}\eta^b l^c(\nabla_c \omega_{b}-\nabla_b \omega_{c})\eqh R_{ab}l^a \eta^b \eqh 0,\label{app:line2_eq2}
\eea
where the second equality is a result of Damour-Navier-Stokes equation (Eq. (6.14) in \cite{Gourgoulhon:2005ng}) on $\H$. Equations \eqref{app:line2_eq1} and \eqref{app:line2_eq2} imply
\beq
u^{(2)}_b \equivh 0.
\eeq

\item Defining $u^{(3)}_b =T^{cd}\left( \omega_b \nabla_c l_{d}- \omega_c \nabla_b l_{d}\right)$ and substituting $\nabla_{b} l_{d}$ from Eq. \eqref{lemma3:eq2}, after straightforward calculations one gets
\beq
u^{(3)}_b \equivh 0.
\eeq

\item $u^{(4)}_b = T^{cd}\left(\nabla_c l_{b}-\nabla_b l_{c} \right)k^e \nabla_e l_{d} \equivh 0$ by substituting $\nabla_c l_{b}-\nabla_b l_{c}$ from Eq. \eqref{lemma3:eq2} and carrying out the calculations.

\item Lastly, we define $u^{(5)}_b=-T^{cd}\left(l_b \nabla_c -l_c \nabla_b \right)(k^e \nabla_e l_{d})$. We get the desired result as follows
\bea
u^{(5)}_b && \equivh l_c T^{cd} \nabla_b (k^e \nabla_e l_{d})\notag\\
&&\eqh \mathcal{T} l^d \nabla_b (k^e \nabla_e l_{d})\notag\\
&&=\mathcal{T} \nabla_b (l^d k^e \nabla_e l_{d}) - \mathcal{T} k^e \left(\nabla_e l_{d}\right)\left(\nabla_b l^d\right)\notag\\
&&\eqh - \mathcal{T} k^e \nabla_e l_{d}(\omega_b l^d -l_b k^c \nabla_c l^d)\notag\\
&&\equivh - \mathcal{T} l^d \left(\nabla_e l_{d}\right) \omega_b k^e=0,
\eea
where we have used $l_a l^a =0$ and $\nabla_b l_{a}l^a =0$.
\end{enumerate}
We have shown that all five terms on the right-hand side of Eq. \eqref{lemma3:eq3} are proportional to $\xi^a$, thus the proof is complete.

\section{Comparison with ref. \cite{Tanahashi:2017kgn}}\label{app:ref_comparison}
Here, we investigate the relation between our result and Ref. \cite{Tanahashi:2017kgn}. In particular, how the principal symbol $P$ is related to the effective metric for linear perturbations $f^{ab}$. To recap, the principal symbol  $P$ of surface $\Sigma$ is defined as
\beq
P = \frac{\partial E_I}{\partial v_{J,ab}}n_a n_b
\eeq
where $E_I$'s are the eoms, $v_J$'s are the degrees of freedom and $n_a$ is the normal vector to $\Sigma$. The key assumption in this definition is that the eoms are quasilinear when expressed in a foliation.

 In Ref. \cite{Tanahashi:2017kgn}   $P$ is calculated without resorting to any perturbative treatment. It is then shown that for stationary spacetime metrics and scalar fields $P$ becomes degenerate on the horizon $\H$ and this is  interpreted to mean that the horizon is a characteristic surface. Taking the stationary limit is rather subtle as, when one imposes stationarity for the whole configuration the equations are no longer hyperbolic and it becomes meaningless to refer to characteristics. One should rather impose stationarity on a background configuration and consider the principal symbol for linear perturbations, which actually coincides with $P$ in a  quasilinear theory. Alternatively, as done in Ref. \cite{Tanahashi:2017kgn}, one could consider a potentially nonlinear perturbation that is localised to the interior of the horizon while taking the stationary limit of the exterior.

In this paper, we take the background metric to be fixed, thus the only degree of freedom is the scalar field. We are  interested in the properties of $P$ on the Killing horizon, so we set $n_a = \xi_a$.
In our setup, this amounts to
\beq
P = \frac{\partial E_\Phi}{\partial \Phi_{ab}}\xi_a \xi_b \eqh 0,
\eeq
where $E_\Phi$ is the scalar field eom. We use the notation introduced in Appendix \ref{appendix_effective_metric} for simplicity, namely $\varphi_a = \nabla_a \varphi$ and $\varphi_{ab} = \nabla_b \nabla_a \varphi$.

However, $\frac{\partial E_\Phi}{\partial \Phi_{ab}}$ is nothing but the effective metric $f^{ab}$. By definition, $f^{ab}$ is the coefficient of the second order field derivative in perturbation around a background configuration. Taking $E_\Phi = E_\Phi(\Phi_a, \Phi_{ab})$, perturbation $\pi$ around a background configuration $\varphi$ results into
\beq
E_\Phi\left(\varphi_a + \pi_a, \varphi_{ab}+\pi_{ab}\right) = E_\Phi\left(\varphi_a, \varphi_{ab}\right) + \frac{\partial E_\Phi}{\partial \Phi_{a}} \pi_a +\frac{\partial E_\Phi}{\partial \Phi_{ab}} \pi_{ab},
\eeq
which gives
\beq
f^{ab} = \frac{\partial E_\Phi}{\partial \Phi_{ab}}.
\eeq
Consequently, the result of Ref. \cite{Tanahashi:2017kgn} can be translated to our notation as
\beq
f^{ab} \xi_a \xi_b \eqh 0
\eeq
which is Eq. \eqref{cond1}.

\end{document}